\documentclass[journal]{IEEEtran}

\usepackage{graphicx} 
\usepackage{tikz}

\usepackage{amsmath}
\usepackage{xurl}
\usepackage[breaklinks=true, hidelinks]{hyperref}

\usepackage{xcolor}
\usepackage{color, colortbl} 
\usepackage{algorithm,multicol}
\usepackage{algpseudocode}

\usepackage{tabularx}
\usepackage{draftwatermark}
\usepackage{booktabs}

\begin{document}
\SetWatermarkText{PREPRINT}
\SetWatermarkScale{1}
\SetWatermarkLightness{0.98}

% \title{An Emulation-Anchored Digital Twin for Hospital IT–OT Testbed for Cyberattack Analysis}
\title{An Emulation-Anchored Digital Twin Testbed for Cyberattack and Defense Analysis in Hospital IT–OT Environments}

\author{
Prashant Rawat,
Ravi Kumar Bairagi,
Arunima,
Geeta Yadav
\thanks{
Prashant Rawat, Ravi Kumar Bairagi, Arunima and Geeta Yadav are with the Indian Institute of Technology Ropar, Rupnagar, Punjab, India.

Corresponding author: Geeta Yadav (geeta@cse.iitd.ac.in).
}
}
% \fntext[equal]{Equal contribution.}
\maketitle

\begin{abstract}
Modern hospitals increasingly rely on integrated Information Technology (IT) and Operational Technology (OT) infrastructures to support critical healthcare services. However, this convergence expands the cybersecurity attack surface and makes safe validation of defensive mechanisms difficult on live systems. Existing testbeds often focus on isolated IT or OT environments and do not capture realistic cross-domain healthcare interactions. This work presents a hospital IT–OT cybersecurity testbed coupled with a digital twin for monitoring, experimentation, and validation of countermeasures. The testbed emulates a central server, Electronic Health Record (EHR) systems, SCADA-based infrastructure, and segmented IT, OT, and DMZ networks. It supports controlled cyberattack execution, software-patch evaluation, and training of RL-based defense agents. The testbed is further extended to a digital twin that models the real-time state of the environment using log and network statistics and enables bidirectional interaction through command execution and container lifecycle orchestration. Modbus/TCP and FHIR/HL7 support realistic communication across healthcare and industrial components. Experimental evaluation shows low computation overhead, with average normalized CPU utilization below 0.4\% per container and most lightweight services operating below 0.01\%. OpenPLC Modbus TCP operations achieve a median round-trip latency of 0.901 ms. The testbed also captures a multi-stage SSH-based attack propagating from the DMZ to the IT and PLC networks. The framework provides a foundation for extending the emulated environment toward a hardware-enabled hospital digital twin.

\end{abstract} 

% \begin{keyword}
% \texttt{IT–OT Convergence, ICS Security, Security Testbed, Cyber-Physical Systems, Digital Twin}
% \MSC[2010] 00-01\sep  99-00
% \end{keyword}

\section{Introduction}
Cyberattacks targeting Operational Technology (OT) systems have emerged as a major threat to public services and critical infrastructure \cite{8073874}. Modern OT environments are exposed to a broad range of attacks, including Man-in-the-Middle (MITM) attacks, command injection, sensor spoofing, replay attacks, Denial-of-Service (DoS) attacks, firmware tampering, and Programmable Logic Controller (PLC) logic manipulation \cite{plc}. In hospital environments, such attacks can disrupt life-critical services, including power distribution, water supply, and Heating, Ventilation, and Air Conditioning (HVAC) systems. These services are commonly monitored and controlled through Supervisory Control and Data Acquisition (SCADA) platforms comprising Master Terminal Units (MTUs), Remote Terminal Units (RTUs), PLCs, and Human-Machine Interfaces (HMIs) \cite{OV20241518}.

In parallel, hospitals rely extensively on Information Technology (IT) systems, such as servers, databases, EHR systems \cite{ehr}, and Personal Health Gateways (PHGs). Communication among healthcare systems and medical devices commonly uses standardized protocols such as FHIR/HL7 \cite{6627810}. Although IT--OT convergence improves connectivity and operational efficiency, it also introduces cross-domain cyberattacks. A compromised Internet-facing or enterprise IT service may provide an entry point for lateral movement toward OT networks and, consequently, life-critical hospital infrastructure.

Vendors regularly release software patches to address functional issues and security vulnerabilities in IT and OT products. However, deploying patches, firewall rules, or artificial intelligence-generated mitigation strategies directly on operational hospital infrastructure can disrupt essential medical services \cite{10.1145/3372224.3418162}. Fully autonomous cyber-defense mechanisms may also introduce unintended consequences when they alter network access policies or isolate operational components without adequate contextual awareness \cite{9548018}. Therefore, healthcare cybersecurity requires controlled environments in which defensive strategies can be evaluated safely before deployment and Artificial Intelligence (AI)-generated recommendations can be reviewed by human operators \cite{who2021aihealth}.

This need is aligned with the broader transition from fully automated systems toward human--AI symbiosis in industrial decision-making. In safety-critical environments, AI should augment rather than replace human expertise. AI agents can rapidly analyze system states, identify attack propagation patterns, and recommend mitigation actions, while human operators retain supervisory control over decisions that may affect service continuity \cite{rajgopal_agentic_2025, ahmed_autonomous_2025}. Such collaboration is particularly important in hospitals, where an aggressive mitigation action may contain an attack but simultaneously interrupt a critical healthcare service.

Existing cyber-physical testbeds primarily focus on Industrial Control Systems (ICS), particularly power grids, microgrids, and manufacturing environments. These platforms often employ simulation tools and Hardware-in-the-Loop setups to analyze OT-centric attacks \cite{otoole_cyber-physical_2019,park_advanced_2021,KANDASAMY2022108061}. Although such approaches provide high-fidelity experimentation, they are often difficult to reconfigure and remain largely limited to specific industrial domains. Furthermore, many existing platforms do not adequately represent the interactions among hospital IT systems, healthcare services, and SCADA-controlled infrastructure.

Effective cybersecurity evaluation requires continuous monitoring and command-and-control capabilities to observe system behavior, assess attack impact, and evaluate defensive actions without disrupting the experimental setup. Digital-twin capabilities provide a useful foundation for this purpose by enabling dynamic system-state representation, monitoring, and controlled interaction with cyber-physical components. Recent digital-twin research has explored formal modeling and system-level abstractions \cite{HU2023103880}; however, existing solutions remain predominantly manufacturing-oriented and often lack a security-driven and human-centric focus \cite{DING2026104398}. 
This work addresses these gaps by developing an emulation-anchored hospital IT--OT cybersecurity testbed with digital-twin capabilities. The platform provides a reproducible representation of hospital IT and OT components, supports continuous monitoring, and enables controlled interaction through command execution and container lifecycle orchestration. We use the term \textit{emulation-anchored testbed} to avoid implying real-time bidirectional synchronization with a deployed hospital infrastructure. Extending the platform toward live integration with physical devices and operational systems remains part of future work.

The proposed environment supports realistic multi-stage cyberattack scenarios and controlled evaluation of AI-assisted defense strategies. As a proof of concept, a Reinforcement Learning (RL) agent is trained to mitigate SSH-based brute-force attacks by learning firewall-hardening policies over multiple training episodes. The agent demonstrates how AI can support security operators by rapidly identifying appropriate mitigation actions. 
% Within a human--AI collaborative workflow, the generated actions can be presented as recommendations for operator review, validation, and override before deployment in safety-critical environments. This enables the testbed to serve not only as an experimentation platform but also as a foundation for studying trustworthy, risk-aware, and human-supervised industrial cybersecurity decision support. 

 To support this transition, the proposed architecture provides the control and audit mechanisms required for future human-supervised defense. By securely logging events and enabling granular command-and-control, the testbed serves not only as an experimentation platform but also as a foundational infrastructure for studying trustworthy, risk-aware, and human-in-the-loop cybersecurity workflows.

\textit{Contributions:} The main contributions of this work are as follows:
\begin{itemize}
\item Design and implementation of a unified emulation-anchored hospital IT--OT cybersecurity testbed integrating enterprise systems, healthcare services, and SCADA-based industrial components.

\item Development of digital-twin capabilities for continuous monitoring, system-state representation, command execution, and container lifecycle orchestration.

\item Implementation of a realistic multi-stage cross-domain attack scenario demonstrating lateral movement from an Internet-facing network through the IT environment toward PLC-controlled OT infrastructure.

\item Training and evaluation of an RL-based cyber-defense agent that learns firewall-hardening policies to mitigate SSH-based brute-force attacks.

% \item Establishment of a foundation for human-supervised AI-assisted cybersecurity decision support, enabling operators to evaluate, validate, and refine mitigation actions before applying them to safety-critical hospital infrastructure.

\item Development of control and audit mechanisms within the digital twin architecture, establishing the foundational infrastructure required for future human-supervised, AI-assisted cybersecurity decision support.

\end{itemize}
\textit{Paper organization:} In Section \ref{related_work}, we study related work in brief. In Section \ref{testbed_architecture_section}, we discuss Testbed Design, Implementation, and Modeling, followed by the creation and propagation of attack scenarios in the testbed in Section \ref{attack_propagation}. In Section \ref{evaluation}, we discuss the evaluation of the testbed and the digital twin in detail. We discuss the scope and limitations of fidelity details in Section \ref{scope}.
We conclude the paper in Section \ref{conclusion}.
We list the abbreviations used in the paper in Table \ref{tab:abbreviations}.

\begin{table}[!t]
\centering
\caption{List of abbreviations}
\label{tab:abbreviations}
\begin{tabular}{ll}
\toprule
\textbf{Abbreviation} & \textbf{Definition} \\
\midrule

API & Application Programming Interface \\
CPS & Cyber-Physical System \\
CVSS & Common Vulnerability Scoring System \\
DMZ & Demilitarized Zone \\
DQN & Deep Q-Network \\
IDS & Intrusion Detection System \\

JWT & JSON Web Token \\
LogQL & Log Query Language \\
MDP & Markov Decision Process \\
MRI & Magnetic Resonance Imaging \\

PACS & Picture Archiving and Communication System \\
RBAC & Role-Based Access Control \\
RIS & Radiology Information System \\
RL & Reinforcement Learning \\

SOC & Security Operations Center \\
WAF & Web Application Firewall \\
\bottomrule
\end{tabular}
\end{table}

\section{Related work}
\label{related_work}
\subsection{Cyber-Physical Testbeds and Digital-Twin Capabilities}

Cyber-physical testbeds provide controlled environments for analyzing attacks, evaluating defensive mechanisms, and studying the operational consequences of cybersecurity decisions. Existing testbeds have predominantly focused on Industrial Control Systems (ICS), particularly power grids, microgrids, and manufacturing environments. Table~\ref{background-work-comparison-table} summarizes representative studies and compares their domains, architectures, supported attacks, strengths, and limitations.
\begin{table*}[t]
\centering
\caption{Comparison of Cyber Security Focused Testbeds for ICS}
\label{background-work-comparison-table}

\renewcommand{\arraystretch}{1.35}   % vertical padding
\setlength{\tabcolsep}{10pt}         % horizontal padding

{%
\begin{tabular}{
>{\centering\arraybackslash}m{1.5cm}
>{\centering\arraybackslash}m{1.5cm}
>{\centering\arraybackslash}m{1.8cm}
>{\centering\arraybackslash}m{2.8cm}
>{\centering\arraybackslash}m{3.5cm}
>{\centering\arraybackslash}m{2cm}
}
\toprule
\textbf{Author(s)} &
\textbf{Domain} &
\textbf{Testbed Type} &
\textbf{Key Components} &
\textbf{Attacks Studied} &
\textbf{Strengths}  \\
\midrule

O'Toole et al. \cite{otoole_cyber-physical_2019} &
Power Grid (IEEE-14) &
Real-time HIL &
EMS, Network Sim, OPAL-RT, RTU/IEDs, IDS &
False Data Injection, Denial-of-Service &
Scalable design, IDS integration \\
\midrule

Park et al. \cite{park_advanced_2021} &
Distributed Energy &
Hardware + RTDS &
DERMS, Inverters, ESS, Aggregator &
APT chain, penetration tests &
Strong attack modeling; detailed kill chain  \\
\midrule

Wu et al. \cite{wu_cybersecurity_2023} &
Hybrid Microgrid &
Hybrid Simulation + IEDs &
ABB relays, switches, hybrid infra &
GOOSE spoofing, microgrid threats &
Industrial devices; practical demo  \\
\midrule

Oyewumi et al. \cite{oyewumi_isaac_2019} &
Smart Grid &
Large-scale hardware &
RTDS, substations, SEL IEDs, SDN &
Wide-area attacks &
Realistic, scalable multi-substation setup  \\
\midrule

Fritz et al. \cite{fritz_simulation_2019} &
Synchrophasor System &
Simulation (FPGA) &
PMU, PDC, FPGA &
MITM manipulation; phasor tampering &
Novel FPGA model  \\
\midrule

Our Paper &
Hospital IT–OT &
Docker simulation &
MTU/RTU, PLCs, HVAC sim, WAF, PHG, DMZ &
MITM; PLC maniupulation&
IT–OT integration; patch \& WAF testing  \\
\bottomrule

\end{tabular}
}
\end{table*}

O'Toole et al. \cite{otoole_cyber-physical_2019} present a cyber-physical testbed that integrates an energy management system with the OPAL-RT real-time power-system simulator. The platform supports the emulation of grid behavior and enables the evaluation of attacks such as false data injection and Denial-of-Service (DoS). Communication among cyber and physical components relies on protocols such as DNP3 and IEC~61850. By employing an IEEE 14-bus system, the testbed demonstrates the importance of coordinated cyber-physical modeling for attack-impact assessment.
Park et al. \cite{park_advanced_2021} develop a Hardware-in-the-Loop (HIL) testbed for Distributed Energy Resources (DERs). Their platform incorporates physical hardware and Real-Time Digital Simulator (RTDS)-based components to evaluate cyberattack resilience. The study uses a cyber kill-chain model aligned with the MITRE ATT\&CK framework to analyze multi-stage Advanced Persistent Threats (APTs). This work demonstrates the value of hardware-assisted environments for high-fidelity experimentation.

Wu et al. \cite{wu_cybersecurity_2023} further extend hybrid cybersecurity experimentation by integrating industrial hardware with simulated ICS components. Their platform evaluates protocol-level attacks, including GOOSE-message spoofing in an IEC~61850-enabled environment. Protocols such as Modbus and DNP3 facilitate communication across the testbed. The combination of physical and simulated components improves experimental realism and enables detailed analysis of industrial communication vulnerabilities.

Oyewumi et al. \cite{oyewumi_isaac_2019} present the ISAAC testbed, which supports large-scale cyber-physical experimentation across multiple substations. The platform integrates RTDS-based grid simulation, SCADA components, software-defined networking, and wide-area monitoring. Its modular architecture supports scalable experimentation and the evaluation of distributed cyberattacks. However, such hardware-intensive environments are costly to deploy, difficult to reconfigure, and primarily tailored to power-system use cases.

Fritz et al. \cite{fritz_simulation_2019} adopt a simulation-oriented approach for studying attacks on synchrophasor communication systems. They develop an FPGA-based MITM attack model targeting IEEE C37.118 Phasor Measurement Unit (PMU)-to-Phasor Data Concentrator (PDC) streams. Their results illustrate how manipulated phasor measurements may evade conventional error-detection mechanisms. Although this approach reduces dependency on extensive physical infrastructure, it remains focused on a specific protocol and application domain.

% \begin{table*}[t]
% \centering
% \caption{Comparison of cybersecurity-focused testbeds for industrial and cyber-physical systems}
% \label{background-work-comparison-table}
% % Retain the existing table body after applying the revised row provided below.
% \end{table*}

The reviewed platforms offer important capabilities for high-fidelity OT experimentation. Nevertheless, they primarily emphasize industrial or electrical-system behavior and provide limited coverage of enterprise IT services, healthcare applications, and cross-domain attack propagation. Hardware-heavy environments also impose practical constraints related to cost, scalability, repeatability, and rapid reconfiguration. These limitations are particularly significant in healthcare environments, where cybersecurity evaluation must account for the interaction among IT services, medical-data platforms, segmented networks, and SCADA-controlled infrastructure.

Digital-twin research provides a complementary foundation for monitoring and controlled interaction with cyber-physical environments. Tao et al. \cite{TAO2022372} identify four important modeling dimensions for digital twins: geometry, physics, behavior, and rules. They further classify digital twins according to unit-level, system-level, and system-of-systems representations. Such frameworks provide a structured basis for describing complex environments, although much of the existing work remains oriented toward manufacturing applications.

In healthcare, digital twins have been explored for predictive analytics, resource optimization, and data-driven decision support. Jameil et al. \cite{Jameil2024507} integrate real-time sensor data and historical records to support predictive modeling using Multilayer Perceptron (MLP) networks and XGBoost. In a related study, Jameil et al. \cite{Jameil2024507} address latency reduction in cloud-enabled healthcare digital twins through a Pyomo-based optimization model. Their framework uses the Digital Twin Definition Language (DTDL) and twin graphs to represent relationships among sensors, devices, and virtual models.

Suhail et al. \cite{SUHAIL2025101547} propose a methodology for developing Security-Enhancing Digital Twins (SDETs) for CPS. Their framework includes the prerequisite, design-and-engineering, operation-and-maintenance, and end-of-life phases. It also incorporates security risk management practices and distinguishes between replication-oriented and simulation-oriented digital twins. Replication provides higher fidelity but may be impractical in healthcare due to proprietary systems, heterogeneous devices, and operational constraints. In contrast, simulation- and emulation-based approaches offer greater controllability and repeatability.

Formal modeling techniques further support the systematic design of digital twins. Hu et al. \cite{HU2023103880} employ timed colored Petri nets to represent the discrete dynamics of cooperative manipulation systems. Similarly, the framework in \cite{10128007} uses Unified Modeling Language State Machines (UMLSMs) to describe digital-twin states and transitions within a Digital Twin Runtime Environment (DTRE). In healthcare environments, interoperability standards such as FHIR/HL7 \cite{6627810} additionally support structured data exchange among heterogeneous systems.
A human digital twin framework for monitoring and modeling individual health and occupational well-being was developed \cite{LYU2026104463}; however, unlike their human-centric approach, our work develops a healthcare digital twin testbed for cyberattack simulation, attack impact assessment, and safe validation of defensive countermeasures.
Existing work demonstrates the value of cyber-physical testbeds and digital-twin technologies. Current platforms rarely combine hospital IT services, healthcare-data systems, segmented network zones, and SCADA-based OT components within a unified cybersecurity experimentation environment. The proposed emulation-anchored testbed addresses this gap through a containerized and reconfigurable architecture. It enables safe execution of cross-domain attack scenarios, continuous system monitoring, and iterative evaluation of defensive mechanisms. Unlike a fully synchronized digital twin of a deployed hospital, the current platform provides digital-twin capabilities anchored in emulation. This design choice improves reproducibility and safety while establishing a foundation for future integration with physical devices and live operational data.

\subsection{Human-Supervised Reinforcement-Learning-Based Cyber Defense}

Traditional cybersecurity mechanisms often rely on static rules and manually configured response strategies. Although these approaches remain important, they may be insufficient in dynamic IT--OT environments where attacks evolve rapidly and propagate across interconnected systems. RL offers a complementary approach by enabling agents to learn mitigation policies through repeated interaction with simulated or emulated environments. Table~\ref{tab:lit_review_comparison} summarizes representative AI- and RL-based cyber-defense approaches.
\begin{table*}[t]
\centering
\footnotesize
\caption{Comparison of existing RL-agent-based autonomous cyber resilience systems}
\label{tab:lit_review_comparison}
\renewcommand{\arraystretch}{1.35}   % vertical padding
\setlength{\tabcolsep}{10pt}         % horizontal padding
\begin{tabular}{m{2cm}m{2.7cm}m{3.3cm}m{3.3cm}m{3cm}}
\toprule
\textbf{Paper} &
\centering\textbf{Main Approach} &
\centering\textbf{Techniques Used} &
\centering\textbf{Key Contribution} &
\centering\textbf{Limitations}
\tabularnewline
\midrule

Rajgopal et al. \cite{rajgopal_agentic_2025} &
RL-based autonomous cyber defense &
Deep Q-Network, MDP, TensorFlow &
Automated mitigation with faster response time &
Limited contextual reasoning
\\
\midrule

Saeed et al. \cite{saeed_agentic_2025} &
Hybrid multi-agent cybersecurity framework &
Isolation Forest, XGBoost, CNN-LSTM, PPO, Federated Learning &
Integrated detection, prediction, and response with explainability &
Complex system architecture
\\
\midrule

Ahmed et al. \cite{ahmed_autonomous_2025} &
Hierarchical RL with LLM critics &
RL agents, LLM critics, RAG, MITRE ATT\&CK integration &
Context-aware reward shaping and adaptive learning &
High computational requirements
\\
\midrule

Peter et al. \cite{peter_robust_2025} &
RL-based defense for cyber-physical systems &
PPO, Graph Neural Networks, Grid2Op &
Dual-policy RL for resilient infrastructure defense &
Focus is limited to power grid systems
\\
\midrule

Kuri et al. \cite{kuri_agent_2025} &
Multi-agent phishing detection system &
NLP, Transformer models, Honeypots, Threat intelligence feeds &
Collaborative agents for phishing detection and response &
Limited focus on broader cyber attacks
\\
\midrule

Our Agent &
Network statistics-based detection &
Dynamic Firewall hardening &
Detection of SSH-based attacks and their mitigation &
Focused only on SSH-based attacks
\\
\bottomrule

\end{tabular}
\end{table*}
Rajgopal et al. \cite{rajgopal_agentic_2025} propose an agentic AI framework in which RL agents learn defense policies by interacting with dynamic cyber environments. The problem is modeled as an MDP, in which system states represent packet flows, anomalies, and intrusion attempts, and mitigation actions include blocking suspicious IP addresses, isolating compromised nodes, and initiating sandbox analysis. A DQN is used to learn defense policies through an $\epsilon$-greedy exploration strategy.

Saeed et al. \cite{saeed_agentic_2025} introduce a hybrid multi-agent cybersecurity framework combining machine learning, deep learning, and RL. The framework includes agents for anomaly detection, classification, prediction, and automated response. Isolation Forest and XGBoost support anomaly detection and classification, respectively, while a Graph Neural Network (GNN) captures structural and temporal relationships. A CNN--LSTM model combined with federated learning supports privacy-preserving prediction, and a Proximal Policy Optimization (PPO) agent generates adaptive response actions. SHAP-based explainability is incorporated to improve transparency.

Ahmed et al. \cite{ahmed_autonomous_2025} investigate the integration of Large Language Models (LLMs) with RL to improve contextual awareness during agent training. Their framework pairs RL agents with LLM-based critics that evaluate actions and influence reward shaping. Compact agents monitor individual subsystems, while a central LLM maintains system-level awareness. The study also examines Retrieval-Augmented Generation(RAG)-based knowledge sharing and the integration of external cybersecurity knowledge sources such as the MITRE ATT\&CK framework.

Peter et al. \cite{peter_robust_2025} develop an RL-based defense mechanism for power-grid infrastructures. Their framework combines GNNs with PPO in the Grid2Op simulation environment. A dual-policy learning strategy separates normal operating conditions from critical scenarios, including cascading failures and cyberattacks. This approach demonstrates how RL can support adaptive resilience in safety-critical industrial environments, although the evaluation remains specific to power systems.

Kuri et al. \cite{kuri_agent_2025} propose a specialized multi-agent framework for phishing detection and mitigation. Their system combines detection, deception, collaboration, and response agents. Transformer-based models identify semantic anomalies, honeypots support behavioral analysis, and response agents recommend or implement actions such as IP blocking, email quarantine, and proxy updates. The framework illustrates how multiple agents can coordinate across the detection and mitigation lifecycle.

% \begin{table*}[t]
% \centering
% \footnotesize
% \caption{Comparison of AI-assisted and RL-based cyber-defense approaches}
% \label{tab:lit_review_comparison}
% % Retain the existing table body after applying the revised row and terminology changes provided below.
% \end{table*}

The reviewed studies demonstrate that RL-based cyber-defense mechanisms can improve adaptability, response speed, and scalability. Explainable AI methods, contextual reasoning, and multi-agent architectures can further improve the transparency and effectiveness of AI-generated mitigation actions. However, fully autonomous deployment remains challenging in safety-critical environments. A mitigation action that blocks traffic or isolates a service may limit attack propagation, but it may also disrupt essential hospital operations. Therefore, cybersecurity decision-making in healthcare requires a balance between rapid AI-assisted response and human supervisory control.
This requirement motivates a human--AI symbiotic approach in which the RL agent assists security operators rather than unconditionally replacing them. The AI agent can monitor network statistics, identify attack patterns, and generate firewall-hardening actions, while operators retain the authority to validate, modify, override, or approve high-impact changes. Such a workflow combines machine-speed analysis with human contextual judgment and accountability.
The proposed testbed provides a controlled environment for investigating this interaction. As a proof of concept, an RL agent is trained to mitigate SSH-based brute-force attacks by learning dynamic firewall-hardening policies. The current implementation evaluates the learned mitigation actions within an emulated hospital IT--OT environment. It also establishes a foundation for future human-in-the-loop extensions, including operator approval workflows, risk-aware recommendations, explainable action summaries, and audit logs for AI-assisted mitigation decisions.
Overall, existing literature has not sufficiently addressed the combined need for cross-domain hospital IT--OT experimentation, emulation-anchored digital-twin capabilities, RL-based defensive decision support, and human oversight. The proposed framework contributes toward this gap by providing a safe and reconfigurable platform for studying AI-assisted cybersecurity decisions in safety-critical healthcare environments.
\section{Testbed design, implementation, and modeling}
\label{testbed_architecture_section}
\subsection{Testbed State Representation} 
\label{testbed_state}
The testbed is implemented using Docker as the base platform for simulating an integrated hospital IT–OT environment. Docker provides lightweight containerization, more efficient resource usage, lifecycle management, state observability, and easier deployment compared to virtual machine–based approaches.

The testbed is modeled as a directed graph
\[
G=(V,E),
\]
where \(V=\{P_1,P_2,\dots,P_n\}\) is the set of hospital components. Each component is modeled as a node \(P_i\) , and $E \subseteq V \times V$ is the set of directed edges representing the permitted communication links between them.

\subsection{Routing and Traffic Control}
Based on security zoning principles \cite{belev_purdue_2022}, the architecture is divided into four networks: IT, OT, DMZ, and Internet. Routing and network management are implemented using Alpine-based firewall containers configured with \texttt{ip} and \texttt{iptables}. Firewalls and routers maintain the required network topology, ensuring that the IT network is directly connected to the OT network while preventing direct access to OT from the DMZ and the Internet. 
The set of permitted communication links between network components is denoted by E, where \(E\subseteq V\times V\). For each pair of nodes \((P_i,P_j)\), adjacency indicator is defined as follows:
\[
E_{ij}=
\begin{cases}
1, & (P_i,P_j)\in E\\
0, & \text{otherwise}.
\end{cases}
\]

% In addition to architecture,  Hardware components are emulated using Python-based simulators implementing real protocols such as Modbus/TCP. Attack scenarios can be dynamically introduced across different network segments by configuring the attacker machine. 

The testbed integrates component-level realism via Open-source tools to model core hospital services, such as RapidSCADA, OpenEMR, Orthanc, and the OpenPLC runtime environment. 
The security state $S_i(t)$ of component $i$  within the testbed at time $t$ is defined as

\[
S_i(t)=
\begin{cases}
0, & \text{normal}\\
1, & \text{suspicious}\\
2, & \text{compromised}.
\end{cases}
\]

To effectively evaluate the performance of software patches and attack mitigation strategies, a digital twin continuously maintains the testbed's state and provides the user with command-and-control capabilities. Each component of the testbed is represented by a tuple 
\[P_i(t)=\big(S_i(t),Vul_i,h_i(t),Y_i(t)\big),\]

where the elements are defined as follows:
\begin{description}
    \setlength{\itemsep}{0pt}
    \setlength{\parskip}{0pt}
    \item[$S_i(t)$  :] Inferred discrete security state 
    \item[$Vul_i   $   :] Static vulnerability profile and known software exploits.
    % \item[\textcolor{red}{$\rho_i$      :}] \textcolor{red}{Privilege requirement }
    \item[$h_i(t)$   :] security-relevant Loki hit count for $P_i$
    \item[$Y_i(t)$  :] tshark-derived runtime counter vector.
\end{description}

 The vulnerability score ($Vul_i$) of node \(P_i\) is defined as the average CVSS score \cite{nvd_cvss}  of known vulnerabilities  affecting that component:
% \textcolor{red}{
% \[
% Vul_i=\max_{j\in v_i}\frac{CVSS_j}{10}, \qquad Vul_i\in[0,1].
% \]
% }

% \textcolor{red}{Here, $v_i$ is the set of vulnerabilities associated with node \(P_i\).}

% \textcolor{red}{The privilege requirement of node \(P_i\) is denoted by}
% \textcolor{red}{
% \[
% \rho_i\in\{1,2,3\},
% \]
% }
% \textcolor{red}{which represents the minimum level of access required to compromise the node.}

% \textcolor{red}{The privilege levels are defined as}
% \textcolor{red}{
% \[
% \rho_i =
% \begin{cases}
% 1, & \text{public-level access}\\
% 2, & \text{user-level access}\\
% 3, & \text{administrator-level access}.
% \end{cases}
% \]
% }

% \textcolor{red}{The anomaly score of node \(P_i\) is $L_i(t)$ is set of active security alerts and anomaly logs up to time t}

% \textcolor{red}{and the observed runtime variable of node \(P_i\) is}
% \textcolor{red}{
% \[
% Y_i(t),
% \]
% }
% \textcolor{red}{represent dynamic status of deployed mitigations, patches, or defensive rules.}

The testbed security state vector is
\[
X(t)=\big[S_1(t),S_2(t),\dots,S_n(t)\big]^{\mathsf T}.
\]
\begin{figure*}[t]
  \centering
  \includegraphics[height=9cm,width=1\linewidth]{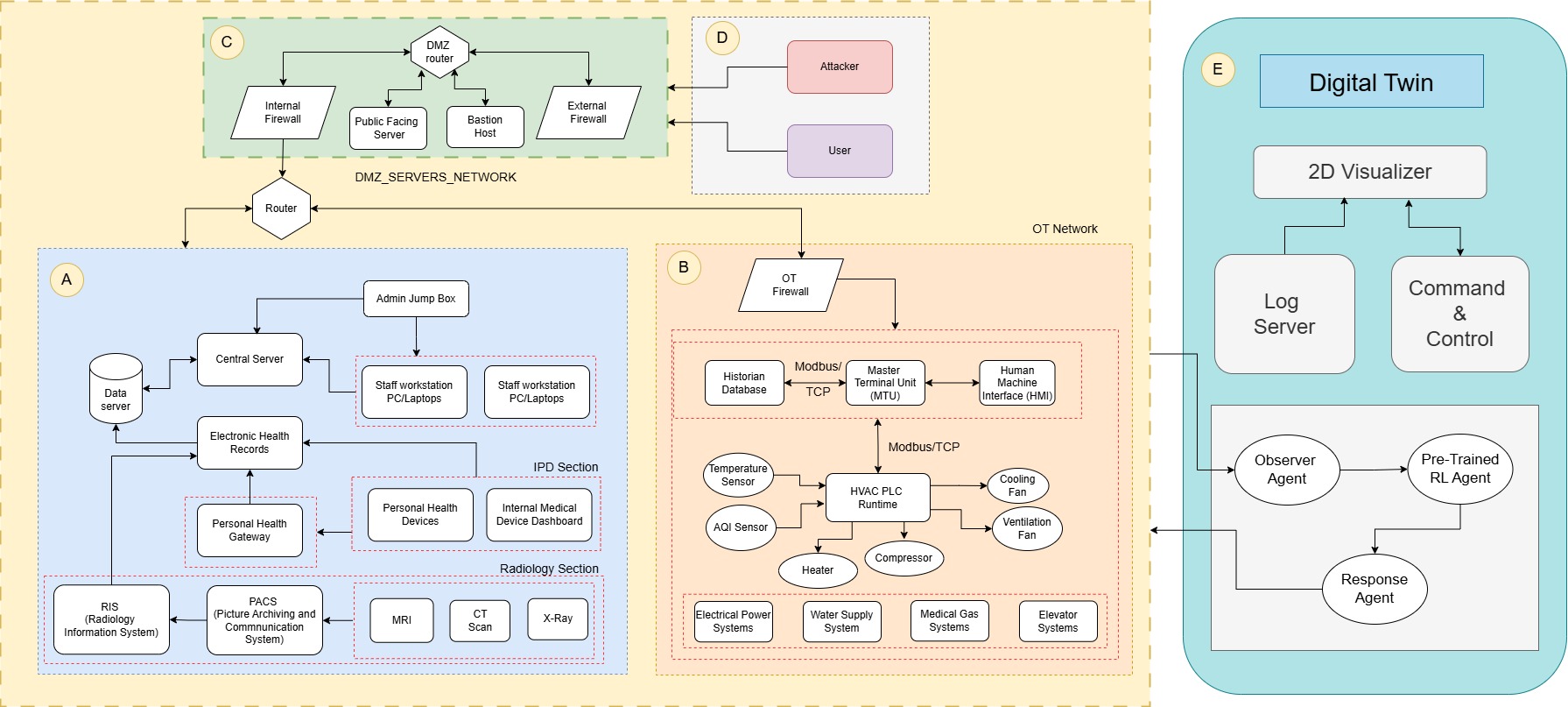}
  \caption{Hospital IT-OT Testbed Architecture. The hospital architecture diagram, with 5 sections (A, B, C, D, and E), depicts the IT, OT, DMZ, Internet, and Digital Twin, respectively.}
  \label{testbed_architecture}
  % \Description{The architecture diagram of the hospital with 3 sections: A, B, and C depicting IT, OT, and DMZ, respectively.}
\end{figure*}

\subsection{Testbed Architecture}
Figure~\ref{testbed_architecture} depicts the proposed hospital architecture, logically divided into five segments: Part A (IT), Part B (OT), Part C (DMZ), Part D (INTERNET), and Part E (Digital Twin). Segmentation facilitates controlled experimentation with various cyberattack scenarios while preserving a realistic hospital environment. Table \ref{system-implementation-mapping} provides a detailed mapping of the software components to their corresponding implementations within the testbed environment.

\renewcommand{\arraystretch}{1.5}
\begin{table}[htbp]
\caption{Mapping of various systems and their implementation}
\begin{center}
\small
\begin{tabular}{cc}
\toprule
\textbf{System} & \textbf{Implementation} \\
\midrule
PLC & OpenPLC Runtime \cite{Alves2014OpenPLC,OpenPLC} \\
\midrule
MTU, HMI, Historian Database & Rapid Scada \cite{RapidSCADA}\\
\midrule
Firewall & ip and iptables rules\cite{Netfilter} \\
\midrule
EHRs & OpenEMR \cite{Purkayastha2019, OpenEMR} \\
\midrule
Sensors and Actuators & Pymodbus (Python) \cite{PyModbus} \\
\midrule
Public Facing Servers & Linux Machine \\
\bottomrule
\end{tabular}
\end{center}
\label{system-implementation-mapping}
\end{table}

\subsubsection{IT Network}
Figure~\ref{testbed_architecture} Part A hosts the core hospital IT infrastructure and is logically separated from the OT and DMZ networks. At its core is a central server connected to a dedicated database for persistent storage and integrated with an Electronic Health Record (EHR) system for medical record management. The EHR is implemented using an OpenEMR instance backed by an external MariaDB container. To ensure controlled administrative operations, access is facilitated through an admin jump box modeled as an Ubuntu container with SSH-based access. Extending beyond core IT services, health-related data components, such as Personal health devices like insulin pumps and pacemakers, are emulated using the pymodbus module that generates realistic event streams. To aggregate the data from these devices, a Personal Health Gateway (PHG) acts as a protocol mediator, normalizing inputs from heterogeneous medical devices and forwarding validated data to patient-specific health monitors using FHIR/HL7 \cite{6627810}. Furthermore, the radiology subsystem includes MRI, CT, and X-ray workflows, with imaging data stored in a PACS server implemented using Orthanc and metadata maintained in a RIS, both integrated with the EHR. The framework supports standard Modbus operations, including read-coil, write-coil, read-register, and write-register transactions. These interactions enable the emulation of HVAC monitoring and control workflows while also supporting cybersecurity experimentation targeting industrial communication channels.
For example, unauthorized Modbus write-coil payloads can be issued to alter HVAC operational states such as fan activation, alarm conditions, or environmental control logic. The resulting operational changes propagate through the SCADA visualization layer, allowing observation of telemetry deviations, actuator-state transitions, and anomalous operational behavior. Rather than attempting hardware-faithful industrial simulation, the framework reproduces protocol-level communication patterns and operational workflows sufficient for cybersecurity experimentation. Consequently, network traces generated within the environment preserve observable communication semantics that can support intrusion detection system (IDS) evaluation, attack-propagation analysis, and anomaly-detection research.

\subsubsection{OT Network}
Figure~\ref{testbed_architecture} Part B represents the OT network and models industrial and SCADA-based systems. It includes a SCADA control station, field devices, sensors, and actuators. An HVAC system is implemented using OpenPLC runtime and Rapid SCADA, enabling real-time monitoring and control of physical processes. Each sensor is accessible through a web-based interface that allows control of its output. The OT environment uses OpenPLC to emulate programmable logic controller (PLC) behavior within the HVAC control workflow. The PLC executes cyclic control logic involving input acquisition, communication handling, logic execution, and output updates, similar to operational PLC workflows used in industrial environments.

Although software-based PLCs cannot reproduce the deterministic hardware timing characteristics of industrial controllers, the framework maintains cyclic control execution behavior suitable for controlled cybersecurity experimentation. This enables the study of attack scenarios involving command injection, unauthorized write-coil operations, register manipulation, and operational state modification within a reproducible environment.
The preservation of scan-cycle behavior is particularly important because many ICS attacks rely on interactions between controllers, SCADA servers, and operational processes. By maintaining operationally meaningful control execution sequences, the framework enables experimentation involving process-state manipulation and attack-propagation analysis across OT infrastructure. The web interface is powered by Python-based backend scripts leveraging the PyModbus library to update the PLC values, thereby generating real OT traffic. In parallel, Rapid SCADA is integrated into the environment to provide core supervisory functionalities, including the Master Terminal Unit (MTU), historian database, and Human–Machine Interface (HMI) for real-time monitoring and control. The traffic is secured by an OT firewall, which allows internal OT traffic and permits network packets from IT. This selective IT and OT interaction is essential for advanced data analytics and software updates. Simultaneously, the firewalls restrict and rate-limit potential propagation of cyber attacks, including PLC manipulation attacks. As a result, the OT environment maintains operational efficiency while preserving critical industrial processes.

\subsubsection{DMZ Network}
Figure~\ref{testbed_architecture} Part C represents the DMZ network, which isolates the IT and OT networks from external entities through a layered firewall architecture. It hosts a dedicated router for its internal networking. The DMZ-external-firewall acts as the public interface, exposing limited services to the Internet, and is responsible for directing incoming and outgoing traffic. DMZ hosts two major services, one of which is a public-facing web server that hosts the hospital's homepage. Another service is SSH-based access to the Admin jump box, provided by the bastion host. It acts as an intermediary node, enabling administrators to access the internal admin jump box for remote management of IT components. On the internal side, a secondary firewall connects the DMZ to a common router, which connects IT, OT, and DMZ with selective traffic control. This layered design enforces strict access control, allowing restricted communication.

% \textcolor{cyan}{
    % \textit{
    The SOC interface mirrors the logical topology of the testbed, grouping the nodes into Internet, DMZ, IT and OT zones. Each node dynamically shows its operational status (normal, warning, alert, offline) based on the status of the container and the indicators derived from the logs. The interface explicitly shows the logging pipeline and the IT–OT routing bridge.
    % }
% }

The containerized architecture establishes the functional foundation required for controlled cybersecurity experimentation. However, evaluating attack propagation and defensive mechanisms requires continuous visibility into component states, system logs, and network behavior. To support this requirement, the testbed is extended with an emulation-anchored digital-twin layer that provides real-time monitoring, controlled interaction, and centralized orchestration.

\subsection{Emulation-Anchored Digital-Twin Layer} 
\begin{figure}[!t]
    \centering
    \includegraphics[width=\linewidth]{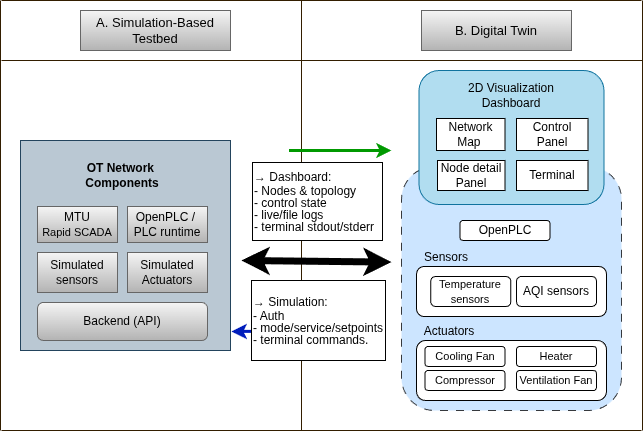}
  \caption{%
(A) Emulation-based testbed—simulated sensors, actuators, PLC, and backend;
(B) Hospital digital twin—2D dashboard for monitoring and control.
}
  \label{testbed_to_digital_twin}
  % \Description{Showing the transformation of the testbed into a digital twin as future work}
\end{figure}

The emulation-based Hospital IT–OT testbed is extended with a closed-loop digital twin that provides an interactive representation of the system (Figure ~\ref{testbed_architecture} part E). The current implementation provides a 2-D interface, as shown in Figure \ref{testbed_to_digital_twin}, that couples the emulated OT plant (PLC, sensors, actuators, and SCADA) with an interactive dashboard for monitoring and control. Key emulated components are interactive, allowing users to inspect real-time operational state and access live system logs. The digital twin continuously ingests system state logs from the testbed and allows the controller to manually execute commands on the testbed components. This feedback mechanism helps to test various countermeasures against cyberattacks.

The digital twin maintains the inferred state vector $X(t)$ using logs, telemetry, and operator feedback from the emulated testbed. The SOC interface renders $S_i(t)$, inferred attack paths, and anomaly scores returned by the propagation engine.

% For each node, the synchronization delay is
% \[
% \Delta_i=t_i^{twin}-t_i^{event},
% \]
% where \(t_i^{event}\) is the event time in the emulated testbed and
% \(t_i^{twin}\) is the time when the event is reflected in the digital twin.

% The bounded-latency condition is
% \[
% \Delta_i\leq \Delta_{\max},
% \]
% where \(\Delta_{\max}\) is selected from measured synchronization latency.

% }

% \textcolor{cyan}{
    % \textit{
    The digital twin is secured with JWT-based authentication and a three-level RBAC model: admin, operator, and viewer. Tokens are time-bound to reduce session exposure. Permissions are enforced at multiple levels: container-level isolation (each role has an allowlist of containers it may access), command allowlisting for remote execution (with dangerous patterns blocked for all roles), and role-specific operational limits such as per-role command timeouts and request rate limiting. This ensures that only authorized actions are performed and reduces the impact of compromised credentials or misuse.
    A secure WebSocket terminal allows authenticated users to run commands inside testbed containers from the digital twin. The endpoint accepts a JWT in the query string and applies the same RBAC rules: access is restricted to containers allowed for the user’s role, and each command is validated against a role-specific allowlist and a global blocklist of dangerous patterns (e.g., shell metacharacters, privilege escalation). Rate limiting is applied per session (e.g., per-role caps on commands per minute). All terminal activity is logged for audit (username, role, container, command, exit code, and result), supporting both real-time monitoring and post-incident analysis.
    % }
    % }
    
 % \textcolor{cyan}{
    % \textit{
    Service management is implemented via the Docker API: start, stop, and restart actions are mapped to the corresponding container lifecycle operations.
    % as shown in Figure~\ref{fig:container_lifecycle_management}. 
    Environmental control allows operators to set temperature and humidity setpoints for logical healthcare zones. Setpoints are stored in the digital twin’s control state and can drive or reflect the behavior of the simulated HVAC and sensors. Network orchestration enforces connectivity between node pairs by dynamically adding or removing iptables FORWARD rules on the IT–OT router container (e.g., block or allow traffic between specified source and destination IPs), enabling experimentation with segmentation and isolation without modifying static firewall configs.
    % }
    % }

\begin{figure}[t]
    \centering
    \includegraphics[width=\linewidth]{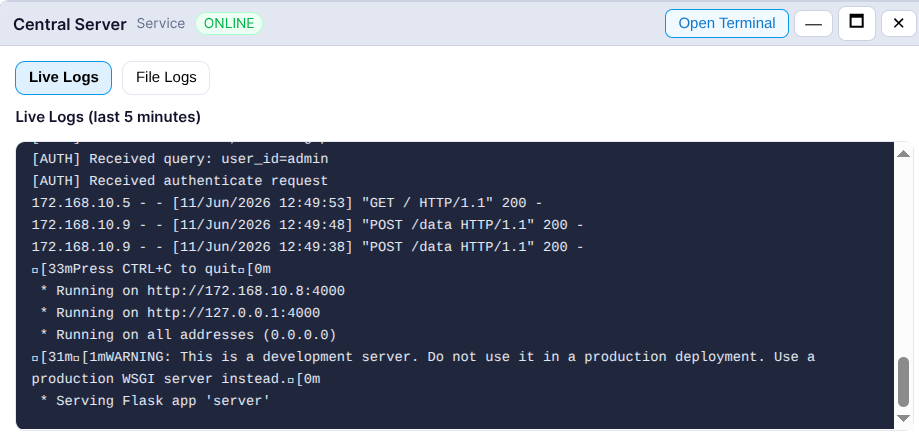}
    \caption{Digital Twin: Runtime container Logs from Docker}
    \label{fig:live-logs-ss}
\end{figure}

\begin{figure}[t]
    \centering
    \includegraphics[width=\linewidth]{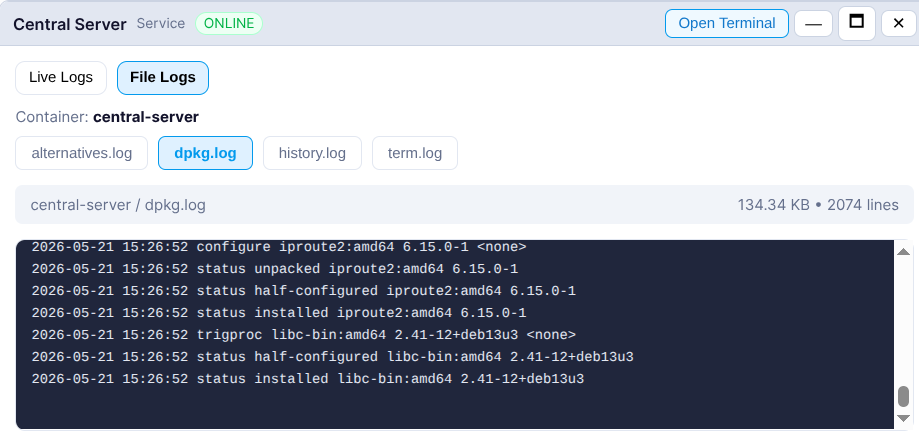}
    \caption{Digital Twin: File-based container-specific logs}
    \label{fig:file-logs-ss}
\end{figure}

 % \textcolor{cyan}{
    % \textit{
   The logging pipeline is implemented as a Loki–Promtail (PLG-style) stack. Promtail discovers running Docker containers, scrapes their logs from the host, and pushes them to Loki with labels such as service, network, and log type as shown in Figure~\ref{fig:live-logs-ss} and \ref{fig:file-logs-ss}. The backend queries Loki’s LogQL API to supply the digital twin with recent logs per node; this powers the real-time diagnostic display in the node-detail view and supports historical audit and security event analysis across the testbed.
   % }
    % }

\begin{figure}[t]
    \centering\
    \includegraphics[width=\linewidth]{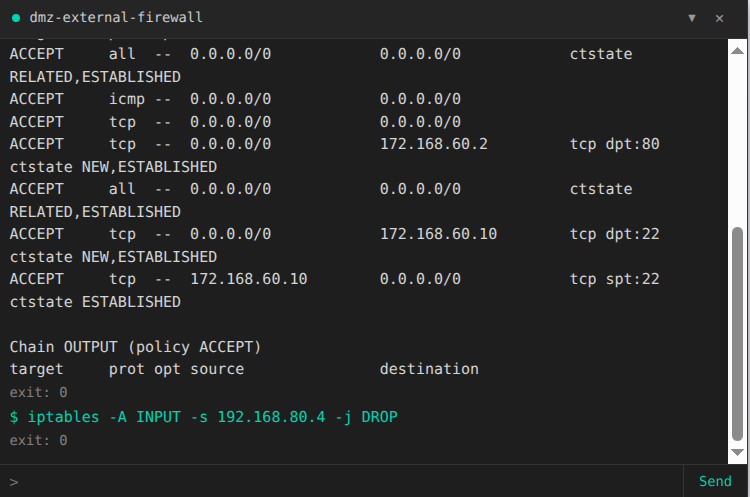}
    \caption{Digital Twin: Command-and-Control interface, configuring firewall rules to block a particular IP.}
    \label{fig:command-and-control}
\end{figure}
\begin{figure*}[t]
    \centering
    \includegraphics[width=\textwidth]{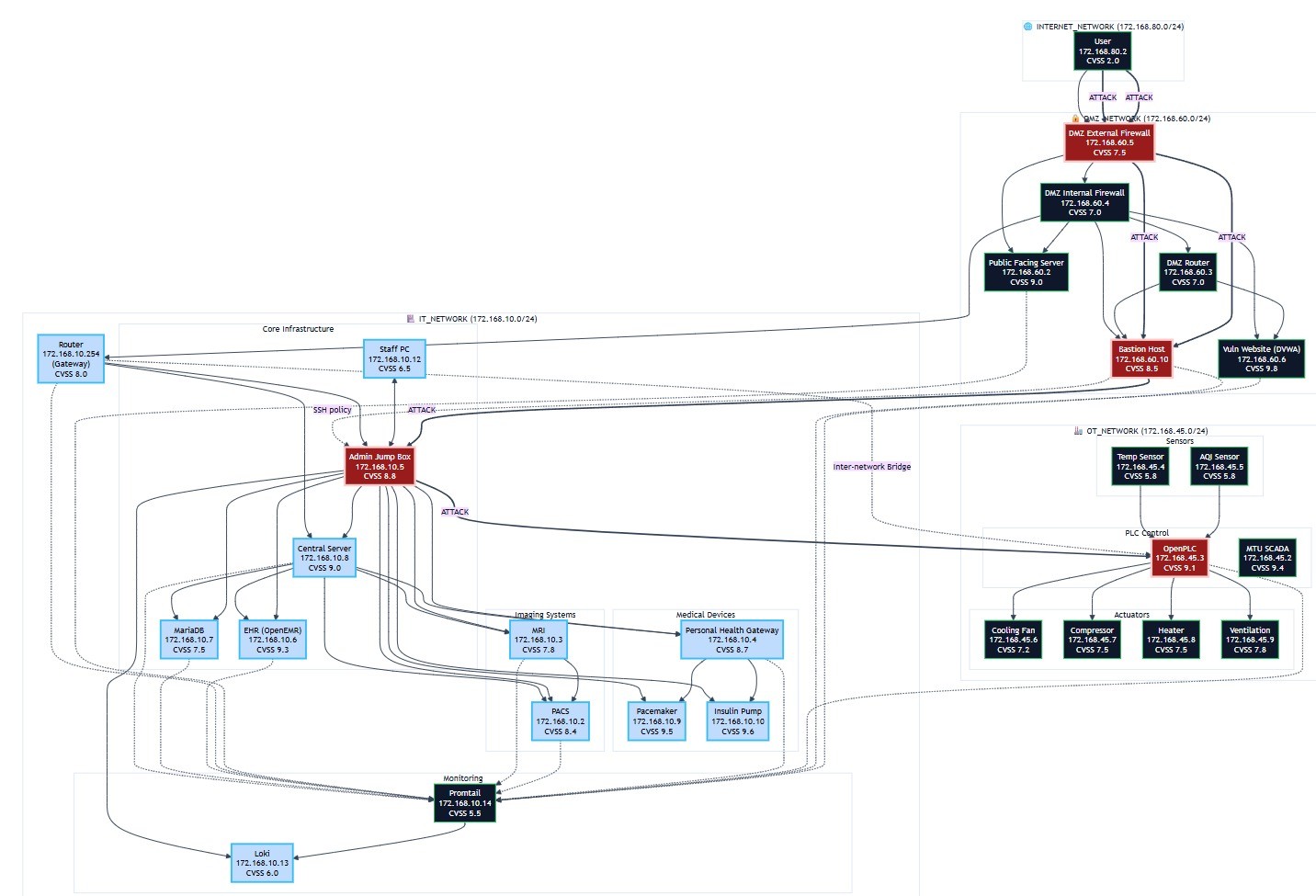}
    \caption{Lateral movement visualization: Simulated attack path across zones. The attacker brute-forces the DMZ bastion, tunnels into the admin jump box, stages a Modbus payload, and pivots to OpenPLC in the OT network. Where the red arrow shows the compromise chain.}
    \label{fig:lateral-movement}
\end{figure*}
 A web-based Command-and-Control (C2) capability is implemented within the testbed to enable centralized management and control of system components. The platform provides real-time system status and the ability to control the component life cycle. A feedback mechanism is provided for executing commands via the Docker command-line interface, as shown in Figure~\ref{fig:command-and-control}. The feedback mechanism allows testing and evaluation of countermeasures against a particular attack scenario. A logging pipeline is engineered to update the control center with the latest system logs for detailed testbed component visualization. 
Together, these access controls and comprehensive logging pipelines form the architectural control and audit mechanisms required to safely implement future human-supervised, AI-driven defense strategies. 
 The objective is to replicate real-world scenarios and determine the optimal solution based on performance metrics.
 The digital twin extends the static hospital topology with a dynamic state that is rendered in the SOC interface through an SOC visualization tab. The tab displays the network as a Mermaid diagram split into zones (INTERNET, DMZ, IT, and OT), where each compromised hop is recorded and overlaid on the diagram in near real-time, as shown in Figure~\ref{fig:lateral-movement}. Analysts can also select any node to highlight it and its direct connections, giving both a live view of propagation across the twin and a simple way to inspect local connectivity.

 % \textcolor{cyan}{
    % \textit{
   % The logging pipeline is implemented as a Loki–Promtail(PLG-style) stack. Promtail discovers running Docker containers, scrapes their logs from the host, and pushes them to Loki with labels such as service, network, and log type. The backend queries Loki’s LogQL API to supply the digital twin with recent logs per node; this powers the real-time diagnostic display in the node-detail view and supports historical audit and security event analysis across the testbed.
   % }
    % }
  The monitoring and control capabilities of the emulation-anchored digital-twin layer enable operators to observe the evolving state of the environment and apply mitigation actions in a controlled manner. Building on this capability, the next subsection introduces an RL-assisted firewall-defense mechanism that learns mitigation policies from network observations and dynamically responds to SSH-based brute-force attacks within the emulated environment.
  
\subsubsection{RL-based Firewall Defense Mechanism}
It is difficult to defend against AI-driven attacks because of their speed and versatility. Existing rule-based defense mechanisms have a limited scope and are static. In contrast, a Reinforcement Learning (RL) agent is adaptive and provides a faster response time. Autonomous decision-making provides real-time defense against attacks \cite{rajgopal_agentic_2025}. 
\begin{figure}
    \centering
    \includegraphics[width=1.0\linewidth]{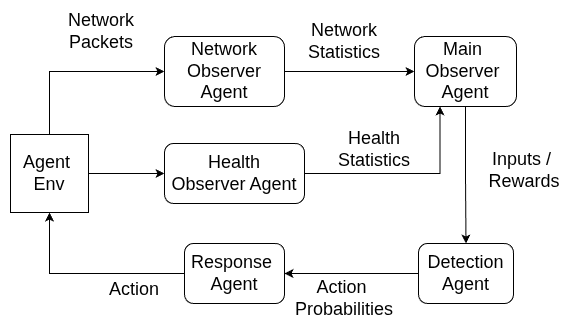}
    \caption{RL Agent Design}
    \label{fig:rl-agent}
\end{figure}
An RL-based Orchestration System is developed that helps the digital twin to autonomously countermeasure the SSH-based attacks. Figure  \ref{fig:rl-agent} illustrates the flow of RL agent training. It consists of MainObserverAgent, NetworkObserverAgent, HealthObserverAgent, DetectionAgent, AgentEnv, Orchestrator, and ResponseAgent. The NetworkObserverAgent collects network packet data from various firewalls present in the testbed and creates statistics for the MainObserverAgent. The MainObserverAgent has a provider method that provides statistics to the DetectionAgent for learning. The ResponseAgent performs the chosen action. The AgentEnv acts as an interface between the testbed and other agents to perform the training of the RL agent. The following subsections describe these elements in detail, beginning with the state and action spaces used by the agent.
\subsubsection{State and Action space}
The primary focus is network-based intrusion detection through firewalls. Tshark is installed on all firewalls and routers to capture the network packets. The NetworkObserverAgent maintains statistics for different firewalls in the form of a dictionary. The source IP, destination port, and protocol are used as keys, and the value is a counter that keeps track of the number of times the particular set of values is encountered. In each step of the training process, the AgentEnv incrementally gets the network statistic for a firewall and performs mapping to integer values to provide input to the model. The output of the model is the expected reward for each action, and the action with the highest reward is passed on to the ResponseAgent. The ResponseAgent receives extra information from the AgentEnv, which includes the source IP and firewall, so that it can perform the following actions on the firewall: Block IP, Block Port, Allow IP, Allow Port, and No Action.
 The DQN architecture provides the learning mechanism required to estimate effective mitigation actions under changing network conditions. To train the agent, the environment repeatedly executes SSH-based attacks and assigns rewards according to the success or failure of the selected defensive actions. The training procedure and reward design are described next.

The detection agent is based on a DQN method \cite{11158143}. It receives input as network statistics based on the protocol and the number of times it is used by a particular IP.  The neural network approximates the Q-function. To remove temporal correlation between samples, a replay buffer is used to store experiences of the RL agent \cite{9758414}. These experiences are randomly sampled and used for training the neural network. A target network is maintained and updated using parameter $\tau$ to stabilize learning.
\begin{equation}
\theta_{target}
\leftarrow
\tau\theta_{online}
+
(1-\tau)\theta_{target}
\end{equation}
For efficient learning, the balance between exploration and exploitation is maintained through an $\epsilon$-greedy strategy. 
\begin{equation}
a_t=
\begin{cases}
\text{random action}, & \text{with probability } \epsilon \\
\\
\arg\max_a Q(s_t,a), & \text{with probability } 1-\epsilon
\end{cases}
\end{equation}
Q-values are updated using the Bellman equation based on rewards from the environment, where 
\(Q(s_t,a_t)\) denotes the action-value function for state \(s_t\) and action \(a_t\), 
\(r_t\) represents the immediate reward received from the environment, 
\(\gamma\) is the discount factor controlling the importance of future rewards, 
and \(\max_{a'} Q(s_{t+1},a')\) denotes the maximum predicted future Q-value for the next state \(s_{t+1}\).
\begin{equation}
Q(s_t,a_t)=r_t+\gamma \max_{a'}Q_{target}(s_{t+1},a')
\end{equation}

\subsubsection{Training and Rewarding the Agent}
\begin{figure*}[!t]
    \centering
    \includegraphics[width=1.0\linewidth]{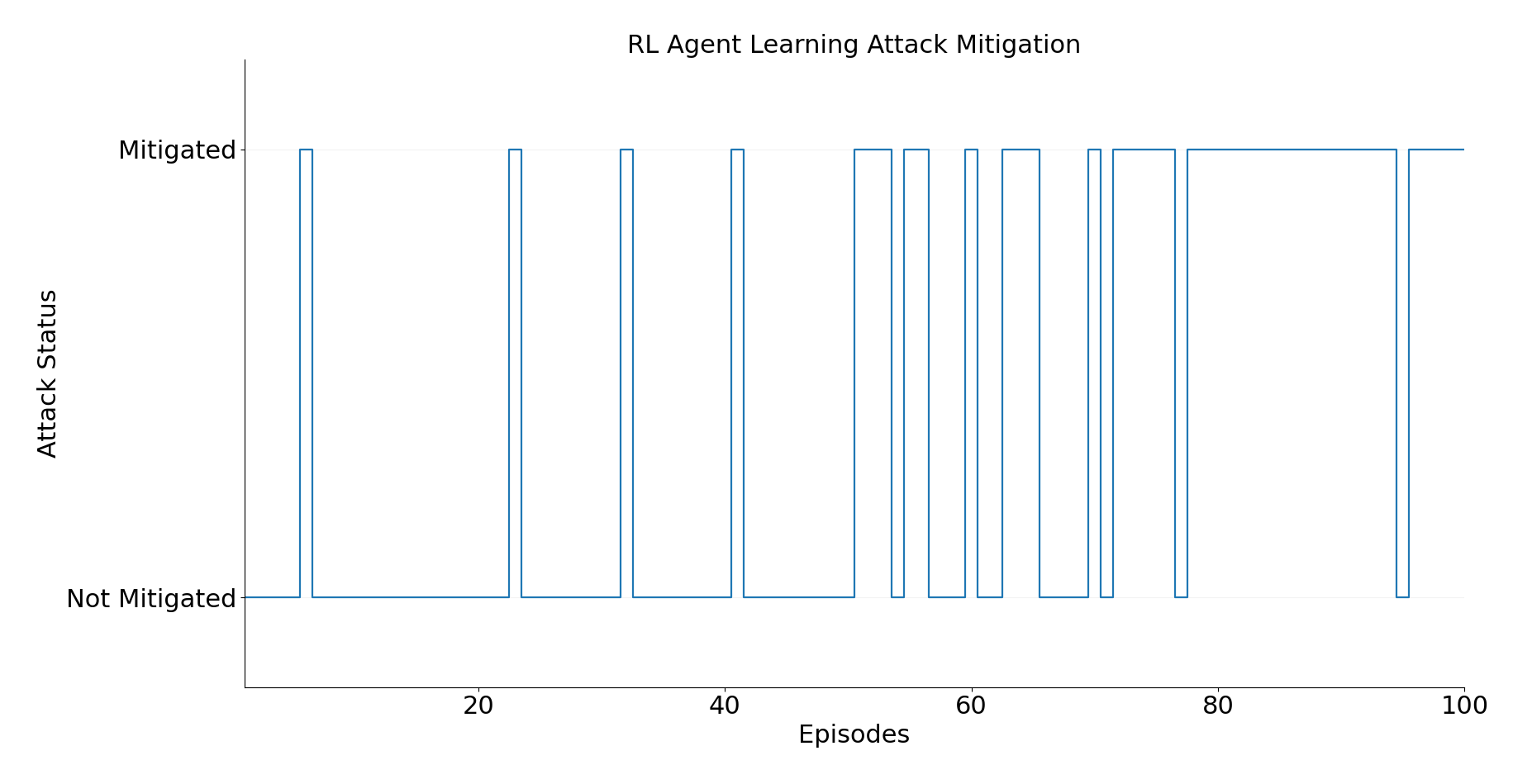}
    \caption{RL agent attack mitigation across training}
    \label{fig:rl-result}
\end{figure*}

The RL agent learns through a series of episodes. Each episode is a new start for the attack. The AgentEnv starts the attack at the beginning of the episode, and when the attack ends or the maximum number of steps is reached, the episode ends. During the episode, the agent performs various actions and receives rewards. The reward mechanism differs from the inputs given to the RL agent. A positive reward of +1 is given when the number of failed SSH attempts on the Bastion Host and Admin Jump Box is zero in the last 5 seconds, which shows that they are not under attack. A negative reward of -1 is given when these two systems are considered under attack. The RL agent begins with exploration and builds logic to understand the consequences of its actions. With the increase in episodes, it moves towards exploitation of its learning and uses it to mitigate the attacks. The RL agent is validated by the mitigation of attacks by the agent, as shown in Figure \ref{fig:rl-result}.
\subsection{Operator Interface and Security Operations View}
\begin{figure*}[t]
    \centering
    \includegraphics[width=\linewidth]{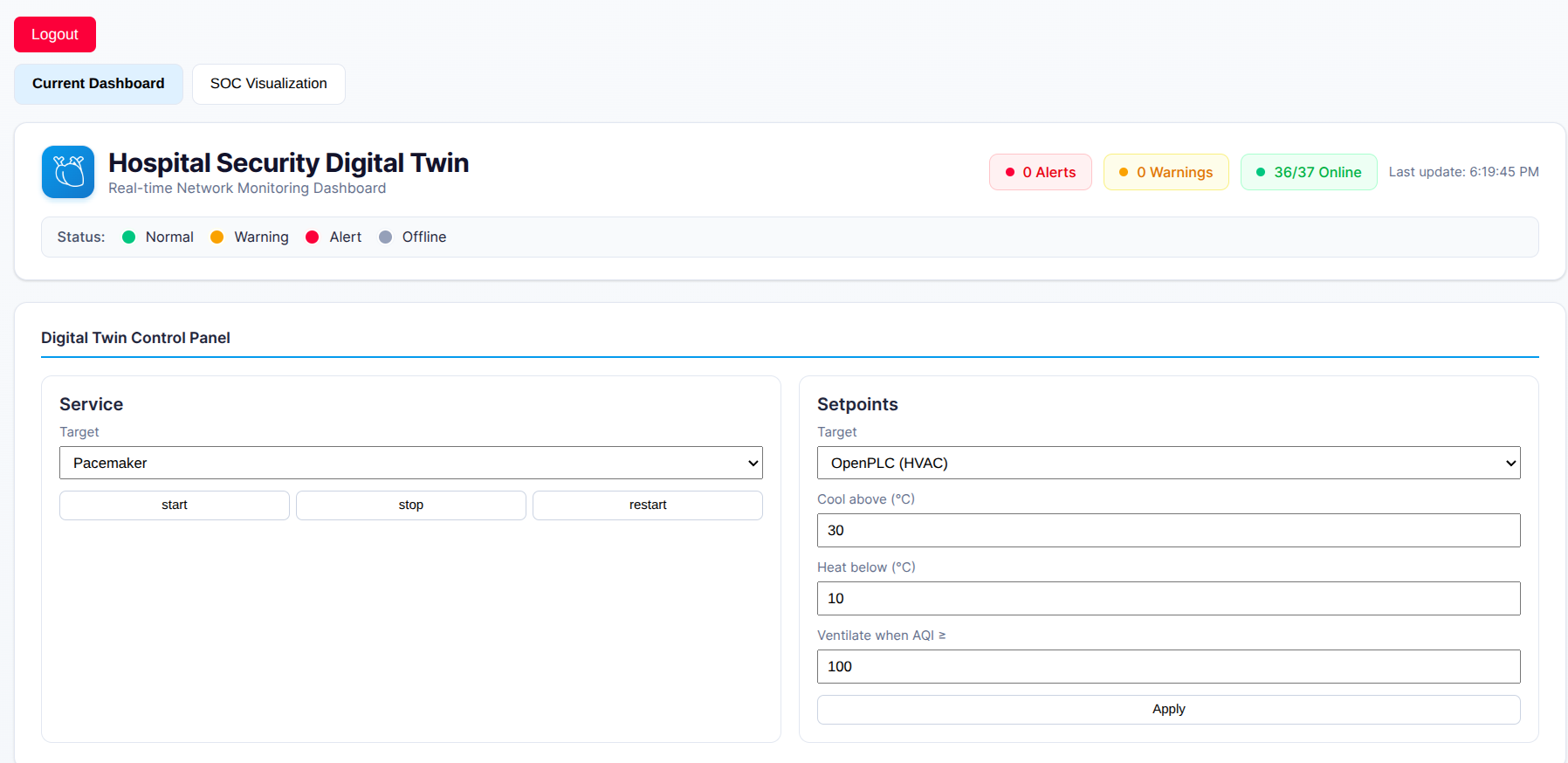}
    \caption{Digital Twin Dashboard: Container Lifecycle Management}
    \label{fig:container_lifecycle_management}
\end{figure*}

Figure \ref{fig:container_lifecycle_management} presents the integrated operator view of the Hospital Security Digital Twin. The main dashboard offers a consolidated health summary, including device online status, alerts, and warnings, and exposes operational controls for representative IT and OT assets. These controls include medical-device lifecycle commands and HVAC setpoint adjustments via OpenPLC.
Complementing the supervisory dashboard, the per-node monitoring panels provide two complementary levels of observability. Live application logs capture runtime behavior in near real time (e.g., authentication events and HTTP traffic on the central server; see Figure \ref{fig:live-logs-ss}), while file-based container logs enable post-hoc inspection of system-level activity (see Figure \ref{fig:file-logs-ss}).
Finally, the SOC visualization maps the segmented network topology—spanning the Internet, DMZ, IT, and OT zones—and overlays lateral movement paths annotated with node-level risk scores. This provides analysts with a spatial perspective on how a compromise could propagate across firewalls, jump hosts, clinical systems, and industrial controllers. Collectively, these views integrate day-to-day simulation control, continuous monitoring, and attack-path reasoning into a unified interface, without duplicating the component-level details presented in the preceding sections.
    
\section{Attack Scenario}
\label{attack_propagation}

% {\color{brown}
\subsection{Attack Propagation}

For successful attack propagation from node $P_i$ to $P_j$, we consider that the attacker has the minimum privilege level required to exploit a vulnerability. For each vulnerability, we calculate the pre-requisite and post-condition based on \cite{9685330}. Thereafter, we consider the possible attack propagation from node $P_i$ to $P_j$ only if the privilege of the attacker at node $P_j$ >= the privilege required to exploit vulnerability $vul$ and communication from \(P_i\) to \(P_j\) is permitted. 

% The attacker is associated with a privilege level

% \[
% \pi(t) \in \{1,2,3\},
% \]

% where \(1\) denotes public-level access, \(2\) denotes user-level access, and \(3\) denotes administrator-level access. Each node \(P_j\) requires a minimum privilege level

% \[
% \rho_j \in \{1,2,3\}
% \]
% /
% to be successfully compromised.

% The impact of the attack influence from node \(P_i\) to node \(P_j\) is defined as a product of 

% \[
% I_{ij}(t)
% =
% \mathbf{1}_{\{S_i(t)=2\}}
% \cdot E_{ij}
% \cdot \mathbf{1}_{\{\pi(t)\ge \rho_j\}}
% \cdot V_j .
% \]

% Here, \(\mathbf{1}_{\{S_i(t)=2\}}\) indicates whether node \(P_i\) is already compromised, \(E_{ij}\) indicates whether communication from \(P_i\) to \(P_j\) is permitted, \(\mathbf{1}_{\{\pi(t)\ge \rho_j\}}\) checks whether the attacker has sufficient privilege to compromise node \(P_j\), and \(V_j\) represents the vulnerability score of node \(P_j\).

% The total attack influence on node \(P_j\) is given by

% \[
% I_j(t)
% =
% \sum_{i:(P_i,P_j)\in E} I_{ij}(t).
% \]

% The compromise probability of node \(P_j\) is modeled as

% \[
% P_j^{\text{comp}}(t)
% =
% 1-e^{-I_j(t)}.
% \]

\paragraph{Direct anomaly estimation}
Each node computes a direct anomaly score from runtime observations.

The log anomaly score is
\begin{equation}
A_i^{\log}(t)=\min\!\left(1,\alpha\, h_i(t)\right)
\end{equation}
where $h_i(t)$ is the security-relevant Loki hit count for $P_i$ in the monitoring window and  $\alpha$ is normalization constant.

The process anomaly score is
\begin{equation}
A_i^{\mathrm{proc}}(t)=\min\!\left(1,\;\sum_{k}\phi_k\!\left(Y_{i,k}(t)\right)\right),
\end{equation}
where $Y_{i,k}(t)$ is the $k$-th \texttt{tshark}-derived counter in $Y_i(t)$, and $\phi_k(\cdot)$ is a piecewise capped scoring function. Representative deployed terms are
\begin{align*}
\phi_{\mathrm{cz}}(y) &= \min(0.35,\,0.08\,y), \\
\phi_{\mathrm{ssh}}(y) &= \min(0.45,\,0.12\,y), \\
\phi_{\mathrm{mod}}(y) &= \min(0.55,\,0.20\,y),
\end{align*}
for cross-zone inbound flows, SSH inbound flows, and non-OT Modbus traffic, respectively. Additional capped terms cover admin-port access, unique peer IPs, and failed SSH logins on bastion/admin nodes. The weights and caps are heuristic calibration constants tuned so benign traffic remains below $\omega_1$, while SSH and IT-origin Modbus events in the implemented attack chain contribute more strongly without saturating the score.

When both signals are available, the fused direct anomaly score is
\begin{equation}
A_i(t)=1-\big(1-A_i^{\log}(t)\big)\big(1-A_i^{\mathrm{proc}}(t)\big).
\end{equation}
If only one signal is non-zero, that score is used directly.

\paragraph{First-pass state estimation} In the first pass, each node is classified using only  log hits $h_i(t)$ and network counters $Y_i(t)$ without considering compromise on neighboring nodes. This yields a preliminary state $S_i^{(1)}(t)$ from the direct anomaly score $A_i(t)$.

Global thresholds are
\[
\omega_1=0.55,
\qquad
\omega_2=0.75,
\]
where $\omega_1$ is the suspicious cutoff (above benign steady-state scores) and $\omega_2$ is sufficiently above $\omega_1$ for confident compromise classification.

The first-pass state is
\[
S_i^{(1)}(t)=
\begin{cases}
2, & A_i(t)\ge\omega_2,\\[4pt]
1, & \omega_1\le A_i(t)<\omega_2,\\[4pt]
0, & A_i(t)<\omega_1.
\end{cases}
\]

\paragraph{Compromise propagation}
Compromised nodes may influence neighbouring nodes through permitted communication links in $G$.

For each directed edge $(P_i,P_j)\in E$, the edge influence term is
\begin{equation}
\label{eq:Iij}
I_{ij}(t)=
\mathbf{1}_{\{S_i^{(1)}(t)=2\}}\,
Vul_j,
\end{equation}
where $\mathbf{1}_{\{\cdot\}}$ is the indicator function,
$S_i^{(1)}(t)=2$ means source node $P_i$ is compromised in the first pass.

The cumulative attack influence on node $P_j$ is
\begin{equation}
\label{eq:Ij}
I_j(t)=\sum_{i:\,(P_i,P_j)\in E} I_{ij}(t),
\end{equation}
and the propagated compromise probability is
\begin{equation}
\label{eq:Pcomp}
p_j^{\mathrm{comp}}(t)=
\begin{cases}
1-e^{-I_j(t)}, & I_j(t)>0,\\[4pt]
0, & \text{otherwise}.
\end{cases}
\end{equation}
\paragraph{Final state estimation}
The final anomaly score and state are
\begin{equation}
A_j^{\mathrm{final}}(t)=1-\big(1-p_j^{\mathrm{comp}}(t)\big)\big(1-A_j(t)\big),
\end{equation}
\[
S_j(t)=
\begin{cases}
2, & A_j^{\mathrm{final}}(t)\ge\omega_2,\\[4pt]
1, & \omega_1\le A_j^{\mathrm{final}}(t)<\omega_2,\\[4pt]
0, & A_j^{\mathrm{final}}(t)<\omega_1.
\end{cases}
\]
The vector $X(t)$ is updated from $\{S_j(t)\}$ at each monitoring interval and displayed in the SOC visualization.

This two-pass procedure first detects locally anomalous nodes, then refines states using graph-based propagation across the hospital IT--OT network.

A patch is considered effective if it reduces the overall anomaly in the system $P_j$:

\[
\sum_{j \in P} A_j^{\mathrm{final}}(t+1)
<
\sum_{j \in P} A_j^{\mathrm{final}}(t).
\]

\subsection{Case Study: Multi-stage SSH Attack}

\paragraph{Baseline Initialization}

We evaluated the propagation model of Section~\ref{attack_propagation} under normal operation before executing attack scenarios.
Constants $\alpha=0.08$, $\omega_1=0.55$, and $\omega_2=0.75$, together with the caps in $\phi_{\mathrm{cz}}(\cdot)$, $\phi_{\mathrm{ssh}}(\cdot)$, and $\phi_{\mathrm{mod}}(\cdot)$, were selected so that benign DMZ web traffic remained below the suspicious threshold.
Table~\ref{tab:baseline-omega} summarizes the resulting baseline scores.

We collected $300$\,s of idle traffic in $56$ snapshots at $5$\,s intervals, with a $30$\,s \texttt{tshark} capture window on firewall/router taps while the full Docker stack ran normally.
Throughout this period, the propagation API reported no inferred lateral-movement path and $S_i(t)=0$ for all monitored nodes.

When benign HTTP traffic was injected (user $\rightarrow$ dmz-external-firewall $\rightarrow$ public-facing-server), only the public-facing server exhibited a stable elevated score, with $A_i^{\mathrm{proc}}(t)=A_i^{\mathrm{final}}(t)=0.35$.
This is consistent with the cross-zone inbound counter $y_i^{\mathrm{cz}}(t)=7$, which yields an uncapped contribution of $0.56$ but is limited to $0.35$ by $\phi_{\mathrm{cz}}(\cdot)$.
Table~\ref{tab:baseline-omega} summarizes mean and maximum final anomaly scores.
All monitored nodes remained below the global suspicious threshold $\omega_1=0.55$,
so $S_i(t)=0$ throughout baseline operation.
Because no node reached $S_i^{(1)}(t)=2$, graph propagation terms satisfied $I_{ij}(t)=0$
and $p_i^{\mathrm{comp}}(t)=0$ for all nodes.

\begin{table}[t]
  \centering
  \caption{Baseline anomaly summary under normal operation (300\,s, 56 snapshots).}
  \label{tab:baseline-omega}
  \begin{tabular}{@{}p{4.5cm}ccc@{}}
    \toprule
    \textbf{Node} $\boldsymbol{P_i}$ & $\boldsymbol{\bar{A}_i^{\mathrm{final}}}$ & $\boldsymbol{\max A_i^{\mathrm{final}}}$ & $\boldsymbol{S_i}$ \\
    \midrule
    Public-facing-server & 0.35 & 0.35 & 0 \\
    \midrule
    User, DMZ-ext.\ firewall, router, staff-pc & 0.00 & 0.00 & 0 \\
    \bottomrule
  \end{tabular}
\end{table}

\begin{figure*}[!t]
    \centering
    \includegraphics[width=1.0\linewidth]{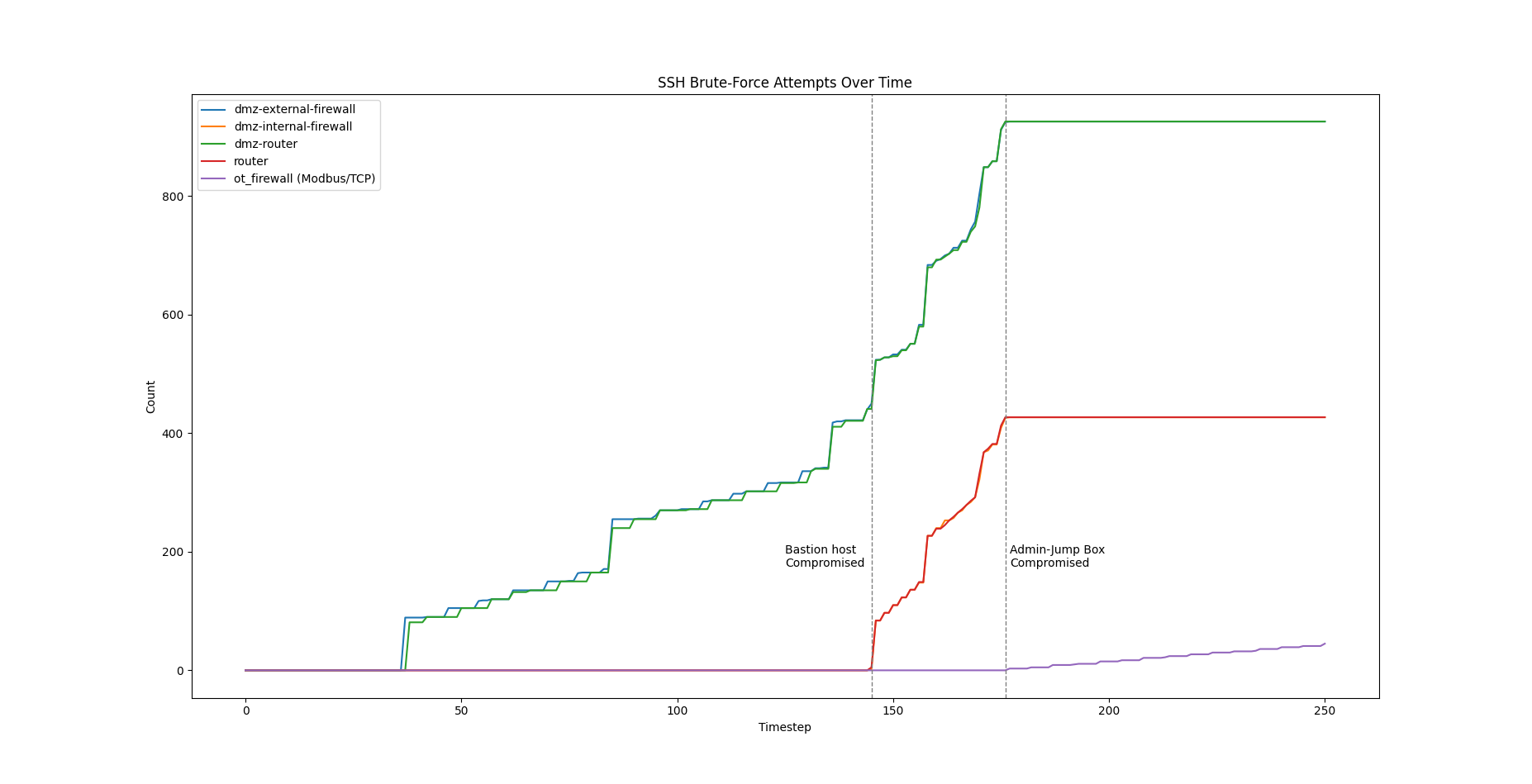}
    \caption{Multi-stage SSH attack propagation. Initially, increased SSH-based traffic is observed in the DMZ-external-firewall and the DMZ-router. After the bastion host is compromised, traffic starts increasing in the DMZ-internal firewall and the main router, indicating lateral movement within networks. Consequently, the OT firewall captures PLC manipulation attacks from IT to OT. }
    \label{fig:attack-propagation}
\end{figure*}

\paragraph{Attack Execution}
A multi-stage lateral movement attack scenario is developed. The attack originates on the Internet and propagates through the DMZ to access the admin jump box within the IT network. The first target is the bastion host in the DMZ network, which is hijacked via a brute-force SSH attack. Once the attacker establishes control of the bastion host, it creates an SSH tunnel for lateral movement to the IT network. As shown in Figure \ref{fig:attack-propagation}, dmz-external-firewall and dmz-router start receiving SSH traffic as the attack starts on the bastion host. Around the 145th timestep, the bastion host is compromised, and the attacker proceeds to the admin jump box via a secure tunnel. The internal DMZ firewall and router also start observing SSH traffic. Around the 178th timestep, the admin jump is compromised, and the attacker has successfully gained control of the IT network. It then starts the attack on the PLC network inside the OT network by using the lack of authentication in the OpenPLC module. The attacker script sends random sensor values to various registers in OpenPLC. The increase in the number of Modbus/TCP calls is observed in the OT firewall from the 178th timestep onwards.

The proposed testbed considers a hospital IT–OT environment consisting of clinical IT services, DMZ-facing gateway components, SCADA infrastructure, PLC-based HVAC control logic, and segmented network zones. The threat model focuses on cyberattacks that originate from compromised IT-facing services or exposed gateway components and propagate toward operational technology systems.
The adversary aims to compromise hospital IT–OT operations by:
disrupting clinical or operational service availability,
manipulating HVAC/SCADA control states,
modifying PLC coil or register values,
bypassing weak segmentation between IT, DMZ, and OT zones,
generating stealthy attack traces for evading detection, and creating cascading effects across healthcare IT and OT workflows. The primary objective is not physical destruction but operational disruption, unsafe control manipulation, and degradation of hospital service continuity.
Adversary Capabilities: We assume the adversary can: scan reachable services across segmented zones, exploit vulnerable IT or DMZ services, obtain a limited foothold on a compromised container or host, perform lateral movement within allowed network paths, issue unauthorized Modbus/TCP commands if OT access is obtained,
manipulate read/write coil or register operations, and observe network traffic visible from compromised nodes. The attacker may exploit misconfigurations, weak credentials, exposed services, insecure communication protocols, or insufficient firewall rules.
Attack Entry Points:
The considered entry points include: \textit{(1) IT zone compromise}, involving the compromise of hospital-facing applications such as EHR, PACS, database services, or administrative workstations; \textit{(2) DMZ compromise}, involving the compromise of gateway components responsible for mediating communication between clinical IT and OT-facing services; \textit{(3) OT network exposure}, resulting from unauthorized access to SCADA, MTU, OpenPLC, or Modbus/TCP endpoints due to weak segmentation or misconfigured routing, or exposed services; and \textit{(4) Remote administrative access abuse}, involving the Misuse of SSH, default credentials, shared accounts, or insufficiently protected jump-box access.
% }

% \subsection{Attack-scenario-2}
% \subsection{Attack-scenario-3}

\section{Evaluation}
\label{evaluation}
The hospital simulation runs as a multi-container Docker stack spanning IT clinical systems (EHR, PACS, MariaDB), OT/ICS components (OpenPLC, MTU, smoke-detector PLC), and DMZ security appliances (firewalls, routers). Before the environment can be reliably used for security experiments, lateral-movement scenarios, or SOC training, its operational viability on commodity hardware must be confirmed. We therefore evaluate per-container CPU and memory consumption during a representative steady-state run. All measurements were collected on a single evaluation host equipped with an AMD Ryzen~5~7535HS processor (6 physical cores, 12 logical CPUs) and 16\, GB of system RAM, running the full Docker stack under Linux.

Docker reports CPU utilization (\texttt{CPUPerc}) as a percentage of a single logical CPU core, which allows individual containers to exceed 100\% on multi-core systems. To express usage as a fraction of the total machine capacity, we normalize the value using the following formula:

\begin{equation}
\text{cpu\_normalized\_percent} = \frac{\text{CPUPerc}}{\text{host\_logical\_cpus}}
\end{equation}
where $host\_logical\_cpus$ is the number of logical processors on the machine that runs the Docker host during evaluation. A logical processor is one schedulable CPU thread exposed to the OS.

Average normalized CPU is the arithmetic mean of all normalized samples over the evaluation window. Peak normalized CPU is the maximum instantaneous value observed.

Memory usage is derived from Docker’s \texttt{MemUsage} \footnote{https://docs.docker.com/reference/cli/docker/container/stats/} field. We report:
\begin{itemize}
    \item Average memory (MB): the mean of \texttt{MemUsed} values, converted to megabytes.
    \item Peak memory (MB): the maximum instantaneous \texttt{MemUsed} recorded during the run.
\end{itemize}

\begin{figure}[htbp]
    \centering
    \includegraphics[width=\linewidth]{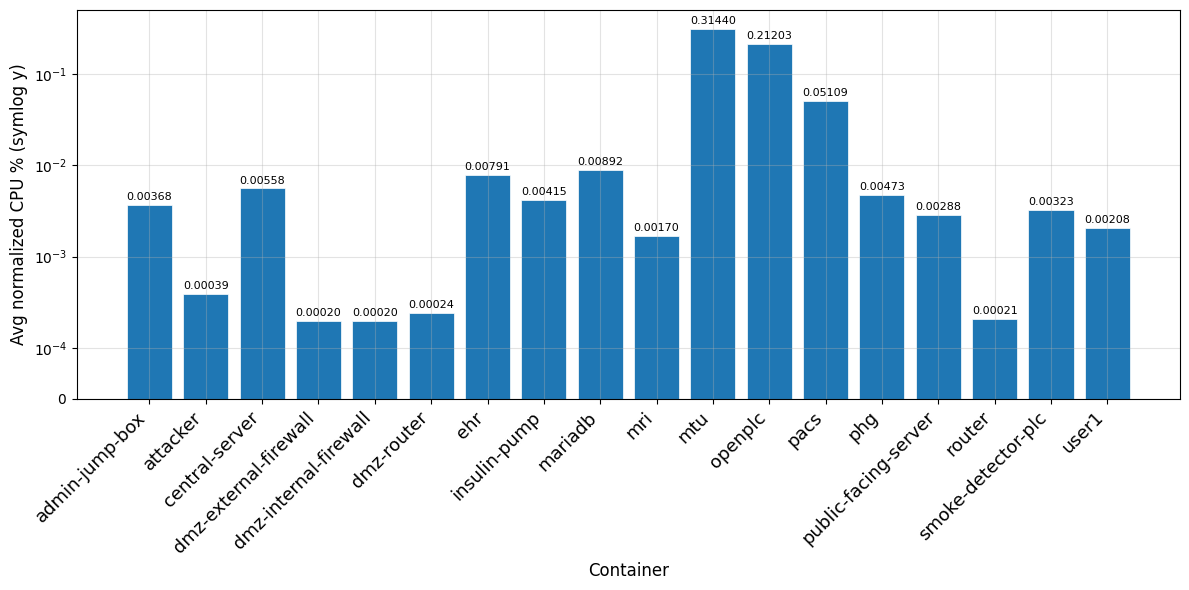}
    \caption{Average normalized CPU utilization per container.}
    \label{fig:avg-cpu-per-container}
\end{figure}

\begin{figure}[htbp]
    \centering
    \includegraphics[width=\linewidth]{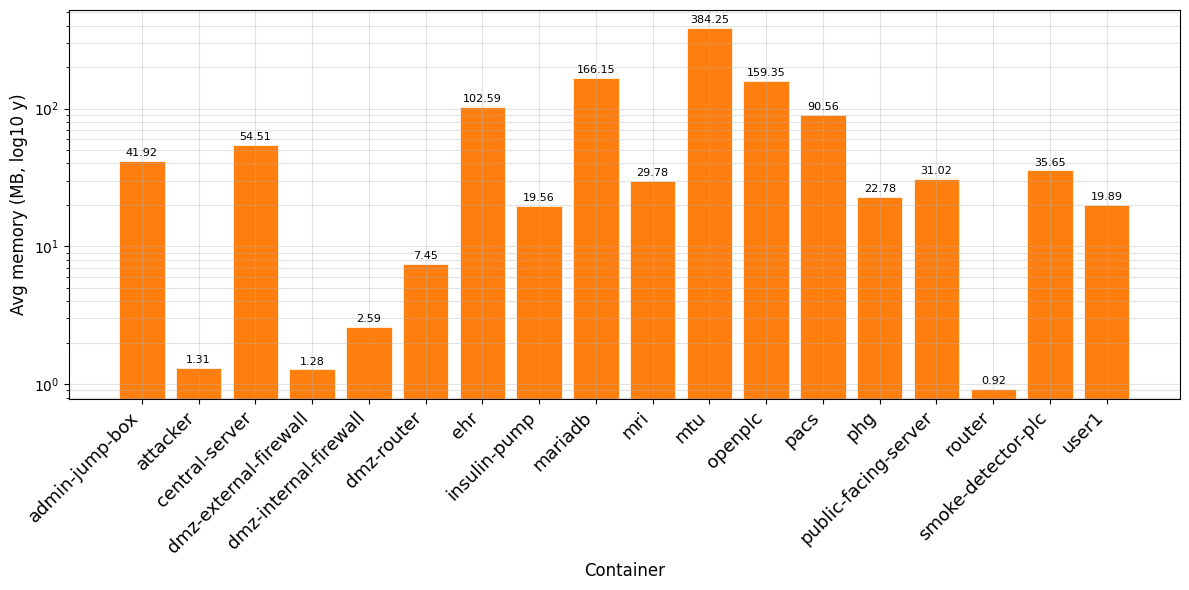}
    \caption{Average memory usage per container.}
    \label{fig:avg-memory-per-container}
\end{figure}

\begin{figure*}[htbp]
    \centering
    \includegraphics[width=\linewidth]{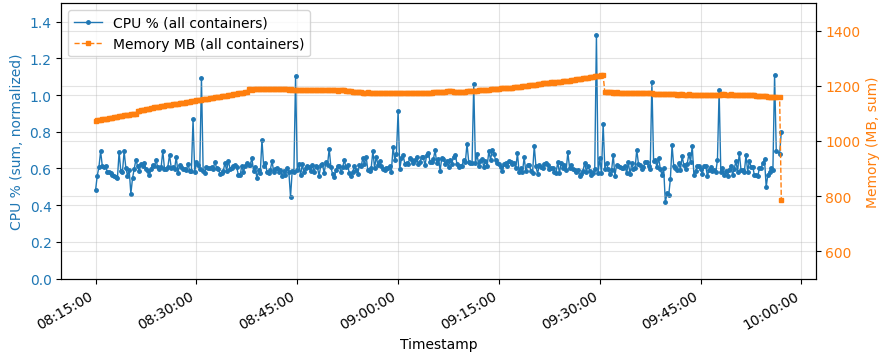}
    \caption{Aggregate CPU and memory utilization of the entire stack over time (1 hour 45 minutes).}
    \label{fig:cpu-memory-timeseries}
\end{figure*}

The simulation demonstrates exceptionally light resource requirements. Average normalized CPU utilization per container remains below 0.4\%, with the majority of services consuming less than 0.01\%. On the 12-logical-CPU host used for evaluation, the full stack represents only a small fraction of available compute capacity.

Notably, the security and networking infrastructure including the router, DMZ external firewall, DMZ internal firewall, and DMZ router, each averages below 0.001\% normalized CPU. This confirms that network segmentation and security policy enforcement impose negligible overhead compared to application workloads.

\begin{table}[htbp]
\centering
\small
\caption{Average Resource Consumption of Major Containers (Steady-State)}
\label{tab:resource-consumption}
\begin{tabular}{p{1.8cm} p{1cm} p{1.5cm} p{2.4cm}}
\toprule
\textbf{Container} & \textbf{Avg CPU} & \textbf{Avg Memory (MB)} & \textbf{Role} \\
\midrule
MTU (Rapid SCADA)     & 0.31\%  & 384 (peak 478) & OT supervisory / HMI layer \\
\midrule
Openplc               & 0.21\%  & 159            & PLC runtime and Modbus services \\
\midrule
PACS (Orthanc)        & 0.05\%  & 91             & Medical imaging archive \\
\midrule
MariaDB               & 0.009\% & 166            & Shared clinical database \\
\bottomrule
\end{tabular}
\end{table}

As shown in Figures ~\ref{fig:avg-cpu-per-container} and ~\ref{fig:avg-memory-per-container}, the MTU is the dominant consumer of both CPU and memory. Running a full .NET-based SCADA server with real-time polling, historian functions, and a web interface, it is substantially more resource-intensive than lightweight network appliances or single-purpose medical device simulators.
OpenPLC ranks as the second-highest CPU consumer, consistent with its role as a soft-PLC runtime that continuously executes structured text logic and serves Modbus/TCP requests on a fixed scan cycle.
MariaDB and PACS (Orthanc) primarily contribute through memory rather than CPU. MariaDB maintains clinical database schemas and connection pools, while Orthanc caches DICOM metadata and imaging buffers. Their low CPU usage indicates they remain largely idle or lightly queried during steady state, yet they reserve a meaningful amount of RAM for operational readiness.

Clinical endpoints (EHR, insulin-pump, MRI, PHG) and supporting IT infrastructure (central-server, public-facing-server, admin-jump-box) form a middle tier, typically consuming 0.002--0.008\% normalized CPU and 20--55 MB of memory.

\subsection{System Stability Over Time}

Figure~\ref{fig:cpu-memory-timeseries} presents the aggregate CPU and memory utilization for the complete 1-hour-45-minute evaluation window with all simulation components active.

The total normalized CPU remains highly stable throughout the run, maintaining a baseline near 0.6\%. Brief periodic spikes reach 1.0--1.1\%, with the highest observed peak at 1.35\%. These excursions are consistent with scheduled SCADA polling cycles and routine background processes rather than uncontrolled load. Overall, aggregate CPU never exceeds 1.4\% of total host capacity.

Memory usage begins at approximately 1,100 MB and rises gradually to 1,250 MB. This modest increase aligns with normal JVM/.NET warm-up and database buffer pool growth rather than a memory leak. A sharp drop to 1,200 MB immediately following the largest CPU spike likely indicates garbage collection or cache flushing in the MTU or MariaDB. At the end of the monitoring window, memory decreases sharply to 800 MB, corresponding to the orderly shutdown of simulation components.

Collectively, the flat CPU baseline, bounded periodic spikes, and controlled memory growth demonstrate that the testbed operates in a stable, steady-state condition. The large margin between observed utilization and host capacity provides ample headroom for concurrent red-team activities, log ingestion, and SOC tooling without compromising system stability.

This lightweight and stable resource profile makes the hospital simulation environment well-suited for extended cybersecurity research and training on commodity hardware.

\subsection{End-to-End Latency Evaluation}

We report latency representing end-to-end round-trip times measured at the client side. Each experiment was repeated multiple times under identical conditions. We characterize the latency distribution of each workload using the following metrics:

\begin{itemize}
    \item \textbf{p50 (median)}: The 50th percentile. Half of all requests are completed in or before this value. It serves as a robust measure of central tendency.
    \item \textbf{p95}: The 95th percentile. Only 5\% of requests are slower than this value. It is used to characterize tail latency while remaining robust against rare outliers.
    
\end{itemize}

\subsubsection{Backend API Latency}

We repeatedly invoked representative backend endpoints and recorded the response time for each request. Table~\ref{tab:backend-api-50-latency} summarizes the results for two key endpoints measured over 50 requests each.

\begin{table}[htbp]
\centering
\caption{Backend API latency (50 requests per endpoint).}
\label{tab:backend-api-50-latency}
\begin{tabularx}{\linewidth}{Xrrrr}
\toprule
\textbf{Endpoint} & \textbf{p50 (ms)} & \textbf{p95 (ms)} & \textbf{min (ms)} & \textbf{max (ms)} \\
\midrule
\texttt{log-structure endpoints} & 2.74 & 3.35 & 2.52 & 3.64 \\
\midrule
\texttt{nodes endpoint}          & 125.30 & 142.43 & 106.05 & 198.95 \\
\bottomrule
\end{tabularx}
\end{table}

The \texttt{log-structure} endpoints return a comparatively small, fixed-size payload and therefore exhibit very low and stable latency. In contrast, the \texttt{nodes} endpoint is substantially slower as it must enumerate and inspect container states, parse detailed metadata, and return a significantly larger result set.

\subsubsection{Log Query Latency}

We evaluated several representative log queries designed to stress different parts of the monitoring stack and query language. Results are based on 20 repeats per query and are presented in Table~\ref{tab:log-query-latency}. To evaluate these results, the queries were categorized into four distinct baselines based on the targeted log streams:

\begin{table}[!t]
\centering
\caption{Log query latency (20 repeats per query).}
\label{tab:log-query-latency}
\begin{tabularx}{\linewidth}{Xrrrr}
\toprule
\textbf{Query} & \textbf{p50 (ms)} & \textbf{p95 (ms)} & \textbf{min (ms)} & \textbf{max (ms)} \\
\midrule
All logs baseline          & 17.48 & 41.45 & 8.99 & 43.78 \\
\midrule
Central server auth signals & 5.58  & 16.96 & 3.48 & 24.21 \\
\midrule
SOC critical services      & 6.43  & 14.57 & 4.75 & 16.64 \\
\midrule
Segmentation firewall signals & 4.80 & 8.40 & 3.50 & 13.63 \\
\bottomrule
\end{tabularx}
\end{table}

\begin{itemize}
    \item \textit{All logs baseline}: A broad log selection constrained to streams with an assigned service identifier, avoiding a full scan of the entire log dataset.
    \item \textit{Central server auth signals}: Authentication-related events originating from the central server.
    \item \textit{SOC critical services}: Restricted to the main IT, DMZ, and SOC services in the scenario.
    \item \textit{Segmentation firewall signals}: Router and firewall logs, with emphasis on denied or blocked traffic.
\end{itemize}

\subsubsection{OpenPLC Modbus TCP Control-Plane Latency}

We measured a lightweight control-plane interaction against OpenPLC over Modbus TCP. The test consisted of writing three holding registers followed by an immediate read-back. Across 20 trials, the measured round-trip latency showed a minimum of 0.753\, ms, a median (p50) of 0.901\, ms, a 95th percentile (p95) of 3.435\, ms, and a maximum observed value of 47.445\, ms.

The occasional high tail latency (maximum of 47.445 ms) is attributed to normal system-level variability such as OS scheduling, TCP stack behavior, and measurement overhead.

This low-latency profile confirms that the simulated industrial control systems remain responsive and suitable for realistic security experimentation and operator training.

\section{Scope and Limitations of Fidelity}
\label{scope}
The proposed framework is designed as a cybersecurity experimentation and attack-analysis environment rather than as a hardware-faithful industrial simulator or a fully synchronized digital twin of a deployed hospital infrastructure. Its objective is to reproduce the communication patterns, segmented network behavior, and operational workflows required to study cybersecurity interactions across hospital IT and OT domains in a safe and reconfigurable setting.
The current implementation emulates protocol-level interactions and observable cyber-physical behavior using containerized services, software-based PLCs, and representative sensors and actuators. However, it does not claim deterministic timing equivalence with physical PLC hardware or deployment-grade industrial control systems. In particular, the present framework does not model hardware-specific PLC execution characteristics, physical sensor noise, electrical or thermodynamic process dynamics, actuator inertia, or vendor-specific firmware behavior.
Instead, the framework prioritizes operationally meaningful communication semantics and security-relevant system behavior. This level of abstraction is sufficient for evaluating cross-domain attack propagation, intrusion-detection mechanisms, software-patch validation, firewall-hardening strategies, and AI-assisted mitigation approaches in hospital IT--OT environments. The architecture also provides a foundation for future extensions involving physical devices, hardware-in-the-loop integration, and real-time synchronization with operational infrastructure.
\section{Conclusion and Future Work}
\label{conclusion}
The proposed framework provides a containerized hospital IT--OT cybersecurity testbed that integrates isolated services, segmented networks, healthcare applications, and SCADA-based OT components within a unified experimentation environment. Although the current implementation relies on software-emulated PLCs, sensors, and actuators, it preserves the protocol-level interactions, communication semantics, and operational workflows required for studying cross-domain attack propagation and defensive mechanisms. The framework is therefore positioned as an emulation-anchored cybersecurity testbed with digital-twin capabilities rather than as a hardware-faithful industrial simulator or a fully synchronized digital twin of a deployed hospital infrastructure.
The testbed enables safe and repeatable evaluation of multi-stage cyberattacks, system monitoring, firewall-hardening strategies, and RL-assisted mitigation mechanisms without disrupting operational healthcare services. As a proof of concept, the RL agent demonstrates the ability to learn firewall-defense policies for mitigating SSH-based brute-force attacks and limiting lateral movement across network segments.
Future work will extend the framework through hardware-in-the-loop integration using physical PLCs, sensors, and actuators, as well as real-time synchronization with operational devices. The RL agent will also be enhanced with a richer observation space, an expanded action set, and support for a broader range of attack scenarios. In addition, human-in-the-loop decision support will be incorporated to allow security operators to review, validate, modify, or override AI-generated mitigation actions before their deployment in safety-critical hospital environments.
%  The current work establishes a foundation for a scalable and extensible future hospital IT–OT testbeds and digital twin environments capable of supporting advanced research, attack modeling, and intelligent defense mechanisms.
\section*{Acknowledgment}
This work was supported by the Anusandhan National Research Foundation ECRG under Grant No. ANRF/ECRG/2024/006514/ENS.
\bibliography{mybibfile}

% \appendix

% \section*{References}

% \bibliography{mybibfile}

\end{document}